\documentstyle[fortschritte,eepic,epic]{article}
 \def\<{\langle} \def\>{\rangle}
\begin{document}
\title{About the use of entanglement in the optical implementation 
of quantum information processing} \author{M. G. A. Paris,  G. M. D'Ariano, 
P. Lo Presti, and P. Perinotti\\ 
Quantum Optics $\&$ Information Group, INFM Udr Pavia, Italy\\ 
{\tt E-mail:paris@unipv.it}, {\tt URL: www.qubit.it/\,$\tilde{}\,$paris}
}\date{}\maketitle
\begin{abstract} We review some applications of entanglement to 
improve quantum measurements and communication, with the main focus 
on the optical implementation of quantum information processing. 
The evolution of continuos variable entangled states in active 
optical fibers is also analyzed.
\end{abstract}
\section{Introduction}\label{s:intro}
Quantum information theory has developed dramatically over the past 
few years, driven by the prospects of quantum-enhanced communication, 
measurements and computation systems. Most of these concepts were 
initially developed for discrete quantum variables, in particular 
quantum bits, which have become the symbol of quantum information 
theory. Recently, however, much attention has been devoted to 
investigating the use of continuous variables (CV) in quantum 
information processing. Continuous-spectrum quantum variables may 
be easier to manipulate than quantum bits in order to perform various 
quantum information processes \cite{polz}. This is the case of Gaussian 
state of light, {\em e.g.} squeezed beams, by means of linear 
optical circuits \cite{pati}. Using CV one may carry out quantum 
teleportation \cite{furu} and quantum error correction. The concepts of quantum 
cloning and entanglement purification \cite{geza} have also been extended to CV, 
and secure quantum communication protocols have been proposed \cite{gran}. \par 
The key ingredients of quantum information is entanglement, 
which has become the essential resource for quantum computing, 
teleportation, and cryptographic protocols. Recently, entanglement 
has been proved as a valuable resource for improving optical resolution 
\cite{fabre}, spectroscopy \cite{spectr}, quantum interferometry
\cite{entdec}, and has shown to be a crucial ingredient 
for making the tomography of a quantum device \cite{tomoch}. \par
In this paper we review some applications of CV entangled states to improve
quantum measurements \cite{entame} and communication in the optical implementation of
quantum information processing. In Section \ref{s:meas} we analyze 
the estimation of a displacing amplitude, also in presence of noise, and 
the case of binary discrimination between two unitary operations.
In Section \ref{s:qint} we study the role of entanglement in improving 
interferometric measurements, and show that an optimized 
two-mode interferometer requires an entangled input state, with 
ultimate scaling that may be achieved using twin-beam in a Mach-Zehnder 
interferometer. In section \ref{s:qcom}, a secret key quantum criptographic 
scheme based on entangled twin beam and heterodyne detection is analyzed and 
shown to be effective both for binary quantum key distribution and as complex 
alphabet trasmission channel. Finally, in Section \ref{s:dent} we study the 
evolution of entangled twin-beam of light in a pair of active optical fibers, 
in order to evaluate the degradation rate of entanglement and determine 
a threshold value for the interaction time, above which the state become 
separable. Section \ref{s:outro} closes the paper summarizing results.  
\section{Entanglement in quantum measurements}\label{s:meas}
The measurement problem we are going to consider is the following:
suppose one is given a quantum device, which perform an unknown  
unitary transformation chosen from a given set, and wants to  
discriminate which transformation (within the set) has been 
actually performed. The unitaries are labelled by a parameter, 
such that the discrimination is equivalent to the estimation of 
the value of the parameter. The inference strategy is that of 
preparing an input probe state and then measuring the outgoing 
signal, such to discriminate among the possible output states. 
In order to achieve the most accurate discrimination one has to 
optimize over the possible input signals and the possible output 
detection schemes. 
The question we want to answer is whether or not entanglement is convenient 
in such discrimination, {\em i.e.} if it is better to use single-mode
probe, or to place the device such to act on a sub\-system of a bipartite 
entangled systems, and then allowing for a measurement on both the modes. 
In the following we consider the estimation of a displacing amplitude, 
also in presence of noise, and the case of binary discrimination,  
{\em i.e.} when our device may perform a transformation chosen from 
a binary set \{$U_1, U_2\}$.
\subsection{Estimation of amplitude}
Let us consider the problem of estimating the amplitude of a displacement
applied to a mode of the radiation field in the phase space, i. e. 
the parameter $\alpha\in{C}$ of the transformation 
$\rho\to\rho_\alpha =D(\alpha)\rho D^\dag(\alpha)$,
where $D(\alpha)=\exp(\alpha a^\dag-\overline{\alpha}a)$.
This transformation can be easily accomplished by a high trasmittivity 
beam splitter and an intense laser beam. For unentangled $\rho$, the 
estimation of $\alpha$ isotropic on the complex plane is equivalent to the
optimal joint measurement of position and momentum, which, 
as well known, is affected by a unavoidable minimum noise of
3dB \cite{yuen82}. Here, the optimal state (for fixed minimum energy)
is the vacuum, and the corresponding conditional probability of
measuring $z$ given $\alpha$ is
$p(z|\alpha)=\pi^{-1}\exp[-|z-\alpha|^2]$. \par
Now, consider the case in
which the estimation is made with $D(\alpha)$ acting on the
entangled state $|x\rangle\!\rangle=\sqrt{1-x^2}\sum_p\: x^p
|p\rangle|p\rangle$, {\em i.e.} the twin-beam state obtained 
by parametric downconversion of the vacuum, with $x\le 1$ (without loss of
generality we may assume $x$ as real) 
and number of photons given by $N=2x^2/(1-x^2)$. 
In this case, the optimal measurement is described by the POVM 
$|z\rangle\!\rangle\langle\!\langle z|$ of
eigenvectors $|z\rangle\!\rangle=D_j(z)\sum_p |p\rangle|p\rangle$ 
($j$ may be either $a$ or $b$) of $Z=a+b^\dag$
with eigenvalue $z$ (this is just a heterodyne
measurement),  now achieving 
$p(z|\alpha)=(\pi\Delta^2_x)^{-1}\exp[-\Delta^{-2}_x|z-\alpha|^2]$, with 
variance $\Delta^2_x=\frac{1-x}{1+x}$ that, in 
principle, can be decreased at will with the state $|x\>\>$
approaching a state an eigenstate of $Z$ (by increasing the gain of
the downconverter). \par  
Remarkably, measurement strategies employing entanglement 
are robust against decoherence induced by noise, {\em i.e.} they 
remain convenient also when the estimation is performed with the channel, 
before and after the unknown transformation, affected by noise. 
Let us reconsider the problem
of estimating the displacement in the case of Gaussian noise, which maps 
states as follows $\rho\to\Gamma_{\overline{n}}(\rho)\doteq\int_{C}
\frac{\mbox{d}^2\gamma}{\pi \overline{n}}\exp[-|\gamma|^2/\overline{n}]
D(\gamma)\rho D^\dag(\gamma)$.
The variance $\overline{n}$ of the noise is usually referred to as
``mean thermal photon number''. The case of Gaussian noise is
simple, since one has the composition law 
$\Gamma_{\overline{n}}\circ\Gamma_{\overline{m}}
=\Gamma_{\overline{n}+\overline{m}}$,
and $\Gamma_{\overline{n}}[D(\alpha)\rho D^\dag(\alpha)]=
D(\alpha)\Gamma_{\overline{n}}(\rho) D^\dag(\alpha)$. Therefore, if the
measurement is made on the entangled state $|x\>\>$ one can easily
derive a Gaussian probability distribution with variance
$\sigma_2^2=\Delta^2_x+2\overline{n}_T$, where $\overline{n}_T$ is the
total Gaussian noise before and after the displacement
$D(\alpha)$, and the noise is doubled since it acts indenpently on
the two entangled beams. On the
other hand, in the measurement scheme with unentangled input (remind
that the optimal probe is the vacuum), one has
$\sigma_1^2=1+\overline{n}_T$. One concludes that the entangled input is
no longer convenient if $\sigma_2^2 < \sigma_1^2$, {\em i.e.} above one 
thermal photon $\overline{n}_T=1$ of
noise. This is exactly the threshold of noise above which the
entanglement is totally degraded to a separable state
\cite{simon}, and therefore the quantum capacity of the
noisy channel vanishes. Since at optical frequencies $\bar{n}_T$ is 
a small quantity we conclude that entanglement is convenient also 
in presence of decoherence induced by noise. 
\subsection{Binary discrimination}
Let us suppose that we have to distinguish among two unitaries $U_1$
and $U_2$. Given an input state $|\psi\>$, one optimizes over the possible
measurements, and the minimum error probability
in discriminating $U_1|\psi\>$ and $U_1|\psi\>$ \cite{helstrom} is given, 
in a Bayesian approach, by
\begin{eqnarray}
P_E={1\over2}
\left[1-\sqrt{1-|\langle\psi|U_2^\dag U_1|\psi\rangle|^2}\right]\;,\label{PE}
\end{eqnarray}
so that one has to minimize the overlap $|\langle\psi|U_2^\dag
U_1|\psi\rangle|$ with a suitable choice of $|\psi\>$. Chosing as a
basis the eigenvectors $\{|j\>\}$  of $U_2^\dag U_1$, and writing
$|\psi\>=\sum_j \psi_j |j\>$, we define $z_\psi\doteq \langle
\psi|U_2^\dag U_1|\psi\rangle = \sum_j |\psi_j|^2 
e^{i \gamma_j}$, where $e^{i \gamma_j}$ are the eigenvalues of $U_2^\dag U_1$. The
normalization condition for $|\psi\>$ is $\sum_j|\psi_j|^2=1$, so that
the subset $K(U^\dag_2 U_1)\subset {C}$ described by $z_\psi$
for varying $|\psi\>$ is the convex polygon having the points
$e^{i\gamma_j}$ as vertices. The minimum overlap 
$r(U^\dag_2U_1)\doteq\min_{||\psi||=1}|
\langle\psi|U_2^\dag U_1|\psi\rangle|$ is the distance 
of $K(U^\dag_2U_1)$ from $z=0$. This geometrical
picture indicates in a simple way what is the best one can do in
discriminating $U_1$ and $U_2$: if $K$ contains the origin then the
two unitaries can be exactly discriminated, otherwise one has to find
the point of $K$ nearest to the origin, and the minimum probability of
error is related to its distance from the origin \cite{entame}. Once the optimal
point in $K$ is found, the optimal states $\psi$ are those corresponding to that
point  through the expression of $P_E$.
If $\Delta(U_2^\dag U_1)$ is the angular spread of the eigenvalues of
$U_2^\dag U_1$ (referring to Fig.1, it is
$\Delta=\gamma_+-\gamma_-$), from Eq. (\ref{PE}) for $\Delta< \pi$
one has $P_{E}=(1-\sqrt{1-\cos^4\frac{\Delta}2})/2$
whereas for $\Delta\ge \pi$ one has $P_E=0$ and the 
discrimination is exact. \vspace{3pt} \\
\begin{minipage}{90mm}
Given a pair $U_1$ and $U_2$ of non exactly discriminable unitaries,
one is interested in understanding whether or not an entangled input 
state could be of some use. The answer is at first negative. In fact, 
using entanglement translates the problem into the one of distinguishing 
between $U_1\otimes I$ and $U_2\otimes I$, thus one has to analyze of 
the polygon $K(U_2^\dag U_1\otimes I)$. Since $U_2^\dag U_1\otimes I$ 
has the same eigenvalues as $U_2^\dag U_1$, the polygons $K(U_2^\dag 
U_1\otimes I)$ and $K(U_2^\dag U_1)$ are exactly the same, so that 
they lead to the same minimum probability of errror.
\end{minipage} 
$\qquad$ 
\begin{minipage}{70mm}
\setlength{\unitlength}{15mm}
\renewcommand{\dashlinestretch}{30}
\begin{picture}(2.5,2.5)(-1.5,-1.5)
\put(0,0){\circle{2}}\put(-1.1,-0.5){\makebox(0,0){\tiny$\gamma_+$}}
\put(0.6,1){\makebox(0,0){\tiny$\gamma_-$}}
\thinlines
\put(0,-1.25){\vector(0,1){2.5}}\put(-1.25,0){\vector(1,0){2.5}}
\path(0.5,0.86)(-0.86,0.5)(-0.94,-0.342)(0.5,0.86)
\texture{8101010 10000000 444444 44000000 11101 11000000 444444 44000000 
        101010 10000000 444444 44000000 10101 1000000 444444 44000000 
        101010 10000000 444444 44000000 11101 11000000 444444 44000000 
        101010 10000000 444444 44000000 10101 1000000 444444 44000000 }
\shade\path(0.5,0.86)(-0.86,0.5)(-0.94,-0.342)(0.5,0.86)
\thicklines
\drawline(0,0)(-0.22,0.26)
\put(-.1,0.15){\tiny$r$}
\put(-1.5,-1.35){\small Fig.1: $r$ is the minimum distance between} 
\put(-0.82,-1.6){\small the origin and the polygon $K$. }
\end{picture}
\end{minipage}  \vspace{5pt} \\ 
The situation changes dramatically if $N$ copies of the
unitary transformation are used (multiple use of the channel). 
In this case one has to compare the ``performance'' of 
$K(U_2^\dag U_1)$ to the one of $K((U_2^\dag U_1)^{\otimes N})$.
Since $\Delta((U_2^\dag U_1)^{\otimes N})=\min\{N\times\Delta(U_2^\dag
U_1),2\pi\}$, it is clear that there will be an $\bar N$ such that
$U_1^{\otimes N}$ and $U_2^{\otimes N}$ will be exactly discriminable, 
{\em i.e.} entanglement makes possible the exact discrimation of {\em any}
pair of unitary transformation. 
\section{Entanglement in quantum interferometry}\label{s:qint}
In this section we want to emphasize the role of entanglement in improving 
interferometric measurements. In particular, we show that an optimized 
two-mode interferometer requires an entangled input state, and that the 
ultimate scaling may be achieved using feasible CV entangled states 
({\em i.e.} twin-beam) in a Mach-Zehnder interferometer. \par 
A general interferometric scheme consists of a source which 
prepares a state $\varrho_0$, an intermediate apparatus which may or 
may not act a perturbation, and a detector described by a generic POVM 
$\Pi$. In a Mach-Zehnder-like interferometer the perturbation is described 
by the unitary operator $U_\phi=\exp\{{\rm i}\phi J_x\}$, where we used 
the Schwinger representation $J_+=a^\dag b\;, J_-=ab^\dag\;,J_z=
\frac12(a^\dag a-b^\dag b)\:,[J_z,J_\pm]=\pm J_\pm\;, [J_+,J_-]=2J_z$ 
in terms of two modes of the interferometer. 
The two possible interferometric outputs are thus given by $\varrho_0$,
if no perturbation occurs, and $\varrho_\phi = U_\phi\:\varrho_0
U^\dag_\phi$, in case of perturbation.  Depending on the outcome of the
measurement one decides for the most probable hypothesis on the state of the
system. Interferometry is thus equivalent to a binary decision problem, and
the corresponding POVM is binary, i.e. the possible outcomes are two
\cite{intbin}.
Optimization of both the input signal and the detection scheme has two main
goals: i) to maximize the probability of revealing a perturbation, when it
occurs, and ii) to minimize the value of the smallest perturbation that can be
effectively detected.  If $\varrho_0$ and $\varrho_\phi$ are orthogonal the 
discrimination is trivial. In general, however, the states are not 
orthogonal and one has to apply an optimization scheme. Since 
interferometric schemes are frequently used for detecting low-rate events, 
we use, rather than Bayesian, the so-called Neyman-Pearson (NP) detection strategy 
\cite{np}, which consists in fixing the false-alarm  probability $Q_0$---the probability
of inferring that the state of the system is $\varrho_\phi$ while it is
actually $\varrho_0$---and then maximizing the detection probability
$Q_\phi$, i.e. the probability of a correct inference of the state
$\varrho_\phi$. The problem is solved by diagonalizing the operator 
$\varrho_\phi - \mu\varrho_0$, $\mu$ (real) playing the role of a 
Lagrange multiplier accounting for the
bound of fixed false alarm probability. The optimal
POVM is the one in which $\Pi_\phi$ is the projection onto the eigenspaces
of $\varrho_\phi - \mu\varrho_0$ relative to positive eigenvalues and
$\Pi_0 ={I}-\Pi_\phi$. If $\varrho_0=|\psi_0\rangle\langle\psi_0|$ and
$\varrho_\phi=|\psi_\phi\rangle\langle\psi_\phi|$ are pure states 
we have $Q_\phi= [\sqrt{Q_0 |\kappa|^2}+\sqrt{(1-Q_0)(1-|\kappa|^2)}]^2$
(if $0\leq Q_0\leq |\kappa|^2$, $Q_\phi$=1, otherwise) 
where $|\kappa|^2=|\langle\psi_0 |\psi_\phi\rangle|^2=|\langle\psi_0
|U_\phi |\psi_0\rangle|^2$ is the overlap between the two states.  
The smaller is the overlap, the easier the discrimination. On the contrary,
when the overlap approaches $1$ one is forced to decrease the detection
probability in order to keep the false alarm probability small.
\par
After having determined the optimal POVM, {\em i.e.} the optimal 
detection scheme, the whole setup can be furtherly optimized looking 
for the best input state, that is a state for which $|\kappa|$ assumes 
its minimum value $|\kappa|_{min}$. As we saw in Section 2.2, the 
value $|\kappa|_{min}$ depends on
the eigenvalues of the unitary operator $U_\phi$. Since 
the spectrum of $J_x$ is the set of relative integers, the
spectrum of $U_\phi$ is the discrete subset $\{{\rm e}^{{\rm
i}m\phi}\;,\;m\in Z\}$ of the unit circle in the complex plane. Apart
from the null measure set $\Phi=\{(q /p)\pi\;,\;q \in 2{Z}+1\;,\;p \in
{Z}\}$ of values of $\phi$, the spectrum of $U_\phi$ is dense in the
unit circle and its convex hull contains the origin of the complex plane,
although there is no couple of diametrically opposed eigenvalues.  If
$\phi\in\Phi$ then the optimal state is given by a superposition of two
eigenstates of $V_{\phi}$ with eigenvalues differing by a factor ${\rm
e}^{{\rm i}\pi}$ \cite{parat}. 
In the general case, the optimal state is any superposition
of three or more eigenstates of $U_{\phi}$, such that the polygon with
vertices on their eigenvalues encloses the origin of the complex plane.
Since $J_x=W^\dag J_z W$ with $W=\exp\left\{{\rm i}\frac\pi 2 J_y\right\}$,
the eigenvectors of $U_\phi$ are entangled.  In fact they are obtained from
the eigenstates of $a^{\dag}a-b^{\dag}b$ (nonclassical states) by the
beam-splitter-like transformation 
$W^\dag=\exp\{-\frac\pi4(a^{\dag}b-ab^{\dag})\}$ \cite{bin}. 
Actually, these optimal
states are far from being practically realizable.  However, we have proved
that they are entangled, and this suggests to explore the possibility of
performing a reliable discrimination by physically realizable entangled
states, e.g. twin-beams $|x\rangle\!\rangle$ \cite{entdec}.
The overlap for the probe prepared in a twin-beam state is given by 
$\kappa=\langle\!\langle x|U_\phi|x\rangle\!\rangle$.
After minor algebra we get 
$$|\kappa|^2=(1+\frac{4x^2\sin^2\phi
}{(1-x^2)^2})^{-1}=[1+N(N+2)\sin^2\phi]^{-1}\:.$$
This value is not zero but it can be arbitrarily small depending on the mean
photon number of the input state. The sensitivity of the interferometer corresponds 
to the minimum detectable value $\phi_{min}$, which is the minimum value of $\phi$ 
such that $Q_\phi/Q_0=\gamma^*\gg1/p$, where $p$ is the {\em a priori} probability 
of the perturbation. The value of $\gamma^*$ is fixed by the experimenter and is 
called {\em acceptance ratio}. In order to understand its meaning we notice that, 
if the setup detects a perturbation, the probability that this inference is true is
$P(p,\phi)=pQ_\phi/[pQ_\phi+(1-p)Q_0]=p \gamma^*/[p\gamma^*+(1-p)]$.
Therefore, the greater is $\gamma^*$, the nearer is this probability to one.
In terms of $|\kappa |$ the condition $Q_\phi/Q_0=\gamma^*\gg1/p$ reads as 
$|\kappa|^2=1 - g(Q_0,\gamma^\star)$ with $g(Q_0,\gamma^\star)=
Q_0 \left[ 1+ \gamma^\star (1-2Q_0) - 2 \sqrt{\gamma^\star 
(1-Q_0)(1-\gamma^\star Q_0)}\right]$
where $|\kappa|^2$ parametrically depends on $\phi$. 
Accordingly, the minimum detectable $\phi$ is given by 
\begin{equation}
\phi_{min}=\arcsin \left(
\sqrt{\frac{\Lambda(Q_0,\gamma^\star)}{1-\Lambda(Q_0,\gamma^*)}}\frac1{
\sqrt{N(N+2)}}\right) \simeq \sqrt{\frac{\Lambda(Q_0,\gamma^\star)}{1-
\Lambda(Q_0,\gamma^*)}}\frac1{N}\:.  \label{minphi}
\end{equation}
Now, we consider twin-beam as input signal of the usual Mach-Zehnder 
interferometer, where the detection stage consists of a difference 
photocurrent measurement. The scheme should be feasible, at least in
principle, and, as we will see, would approach the ultimate
sensitivity bound that has been obtained for the ideal detection. 
After preparation, the twin-beam enters the interferometer, where is 
possibly subjected to the action of the unitary $U_\phi$. 
At the output the two beams are detected and the difference photocurrent 
$D=a^\dag a - b^\dag b$ is measured.  If no perturbation occurs,
then the output state is still a twin-beam, and since $|x\rangle\!\rangle$ is
an eigenstate of $D$ with zero eigenvalue we have a constant zero outcome for
the difference photocurrent. On the other hand, when a perturbation occurs the
output state is no longer an eigenstate of $D$, and we detect fluctuations
which reveals the perturbation.  The false-alarm and the detection 
probabilities are given by $Q_0 = P (d \neq 0 | {\rm not} \: U_\phi) \equiv 0$
and $Q_\phi = P (d \neq 0 |U_\phi) = 1 - P(d\equiv 0 | U_\phi)$, 
where the probability of observing zero counts at the output, 
after the action of $U_\phi$, is given by $P(d\equiv 0 | U_\phi) = 
\sum_n \: \left|\langle\!\langle n,n | U_\phi |x\rangle\!\rangle\right|^2$ 
since the eigenvalue $d=0$ is degenerate.  In this case the false-alarm
probability is zero and therefore it is not necessary to introduce an
acceptance ratio. The scaling of the minimum detectable perturbation can be
obtained directly in term of the detection probability
\begin{eqnarray}
P(d=0 | \phi \neq 0) = 1 - \frac12 \phi^2 N^2 + O(\phi^2)
\quad\longrightarrow\quad \phi_{min} \simeq \frac{\sqrt{2 Q_\phi}}{N}
\label{inttwm}\;.
\end{eqnarray}
One can see that a Mach-Zehnder interferometer fed by twin-beam shows a 
sensitivity that scales with the energy as the ideal scheme.  
Such scaling does
not depend on any parameter but the energy of the input state. This should be
compared with the sensitivity of the customary squeezed states 
interferometry \cite{bon}, where the same scaling is achieved only for a very precise 
tuning of the phase of the squeezing. 
This means that the entanglement-assisted interferometry 
provides a more stable and reliable scheme. 
\section{Entanglement in secure communication}\label{s:qcom}
In this section, a secret key quantum criptographic scheme based on 
entangled twin beam and heterodyne detection is analyzed. The scheme can be 
effectively employed both for binary quantum key distribution and as complex 
alphabet trasmission channel, and the use of entangled signals results in a 
decrease of the error probability. A quantum encoding of the secret-key in 
a cryptographic communication is motivated by the possibility of achieving 
extensive key-expansion, due to the physical limitations in a 
quantum-measurement based eavsdropping. Such an idea for a quantum 
secret-key cryptographic communication was first suggested by Yuen 
\cite{Yuen}.
The secret key is imposed as a random 
displacement transformation, such that the scheme is secure in principle, 
{\em i.e.} the best strategy for an eavsdropper is just pure guess. 
Effects of practical imperfections will be taken into account.
\subsection{Binary communication}
The two value of the bit are encoded in two {\em quasi}-eigenstates of the 
heterodyne photocurrent $Z=a+b^\dag$, {\em i.e.} as 
$"0" \: \rightarrow \: |z_0\rangle\rangle_x = 
D(z_0)|x\rangle\rangle$ and $"1" \: \rightarrow \: |z_1\rangle\rangle_x 
= D(z_1)|x\rangle\rangle $
where $|x\rangle\rangle$ is the twin-beam. 
The $|x\rangle\rangle$'s (and thus the $|z\rangle\rangle_x$'s) become 
orthogonal states for $x\rightarrow 1$. We will use the notation 
$\sigma_0=|z_0\rangle\rangle_x {}_x\langle\langle z_0|$, 
$\sigma_1=|z_1\rangle\rangle_x {}_x\langle\langle z_1|$. \par
The criptographic protocol consists in applying a random displacement
transformation $D(\alpha)$ to the bit {\em before} the transmission.
The value of $\alpha$ represents the key that should be secretely 
shared before the transmission. The receiver (Bob) knows the key, and 
therefore  can apply the inverse transformation $D^\dag(\alpha)$ at the end
of the line and then measure the bit. For this task he has to measure a two-value POVM 
$\{\Pi_0,\Pi_1\equiv1-\Pi_0\}$. The two states are not orthogonal, and 
therefore such a POVM should be optimized to achieve the minimum error 
probability $P_E=\frac12 \hbox{Tr}\left[\Pi_0\sigma_1 + \Pi_1\sigma_0\right]=
\frac12 [1-\sum_i U(\lambda_i) \lambda_i]$, 
where $\lambda_i$ are the eigenvalues of 
the matrix  $\Lambda=\sigma_1-\sigma_0$, and $U(x)$ denotes the Heaviside step function.
Since the two initial states $\sigma_j$ are pure the solution is well known: the POVM is 
projective \cite{helstrom}
(this is true also for mixed $\sigma$'s) and the error probability is given by
$P_E=\frac12 [1-\sqrt{1-\left|{}_x\<\<z_1|z_0\>\>_x\right|^2}]$, where 
$\left|{}_x\<\<z_1|z_0\>\>_x\right|^2=\exp\{-|z_0-z_1|^2 (1+N)\}$. 
For large $N$ or $|z_1-z_0|$ we have $P_E\simeq 1/4 \exp\{-|z_0-z_1|^2 (1+N)\}$.
This result should be compared with the analogue scheme based on displaced
unentangled states {\em i.e.}, $
"0" \: \rightarrow \: |\alpha_0\rangle = D(\alpha_0)|0\rangle$, 
 $"1" \: \rightarrow \: |\alpha_1\rangle = D(\alpha_1)|0\rangle$
where $|\alpha_j\rangle$ are single-mode coherent state and $|0\>$ denotes 
the vacuum state. In this case we have  $P_E=\frac12 [1-\sqrt{1-\left|\<\alpha_1|\alpha_0\>
\right|^2}]$ with $\left|\<\alpha_1|\alpha_0\>\right|^2=\exp\{-|\alpha_0-
\alpha_1|^2\}$ and $P_E\simeq 1/4 \exp\{-|\alpha_0-\alpha_1|^2\}$ for large
$|\alpha_0-\alpha_1|$. As a matter of fact, entanglement is always convenient to 
improve precision of the transmission channel.
\par\noindent 
Let's go back to the entangled scheme: an eavsdropper, say Eve, does not know the key, 
and therefore, to measure the bit, she has (in principle) to discriminate between the 
two mixed states
$\varrho_0 = 1/\pi \int d^2\alpha\: D(\alpha)\: \sigma_0
\:D^\dag(\alpha)$ and  
$\varrho_1 = 1/\pi \int d^2\alpha\: D(\alpha)\: \sigma_1\: 
D^\dag(\alpha)$.
Since the set of displacement operators is a UIR of a 
group we have, according to Schur lemma,
$\Lambda = 1/\pi \int d^2\alpha\: D(\alpha)\:\left(
\sigma_1-\sigma_0\right) \:D^\dag(\alpha) = 
\hbox{tr}\left[\sigma_1-\sigma_0\right] 1 = 0 $, 
and therefore $P_E=1/2$ {\em i.e.} the best strategy for Eve is just pure
guess. 
\par\noindent
In practice, however, it is not possible to impose displacements with uniform
probability in the complex plane. What we can reliably implement is the following
criptographic protocol (random state transformation) 
$\varrho_j =\int d^2\alpha\: g_\kappa(|\alpha|^2)\:  
D(\alpha)\: \sigma_j\: D^\dag(\alpha)$
where $g_\kappa(|\alpha|^2)=\exp(-|\alpha|^2/\kappa)/\kappa\pi$ is a Gaussian
distribution, and to find the error probability for Eve, we have to 
diagonalize $\Lambda = \int d^2\alpha\: g_\kappa(|\alpha|^2)\:  
D(\alpha)\:\left( \sigma_1-\sigma_0\right) \:D^\dag(\alpha)$.
In order to prove that the present protocol is secure we have to compare the
best stratey employable by Eve with a feasible strategy that Bob can use. 
Therefore, we suppose that Eve is trying to eavsdrop a maximally entangled 
channel ($x \rightarrow 1$) which is not perfectly protected ($\kappa$
finite). In this case we have $\varrho_j=D(z_j)\nu D^\dag (z_j)$ with 
$\nu =  \int d^2\alpha\: g_\kappa(|\alpha|^2)\: |\alpha\>\>_1{}_1\<\<\alpha|$,
such that the matrix to be diagonalized is given by  
$$\Lambda =   \int d^2\alpha\: g_\kappa(|\alpha|^2)\:
\left[|\alpha+z_1\>\>_1{}_1\<\<\alpha+z_1| -|\alpha+z_0\>\>_1{}_1\<\<\alpha+z_0| 
\right] = \int d^2\beta \: f(\beta) \: 
|\beta\>\>_1{}_1\<\<\beta |\:,$$ 
with $f(\beta) = g_\kappa(|\beta-z_1|^2)-g_\kappa(|\beta-z_0|^2)$.
The sum $S_+$ of positive eigenvalues of $\Lambda$ correspond to integral 
of $f(\beta)$ over its positivity region {\em i.e.} $|\beta-z_1|^2<
|\beta-z_0|^2$. Suppose that $z_1=a$ and $z_0=-a$, with $a$ real, then 
\begin{eqnarray}
S_+=\int_0^\infty \frac{dx}{\sqrt{\pi \kappa}} 
\left[\exp\{-(x-a)^2/\kappa\}-\exp\{-(x+a)^2/\kappa\}\right] = \hbox{Erf}
\left(\frac{a}{\sqrt{\kappa}}\right)\:.
\end{eqnarray}
The error probability for Eve is thus given by
\begin{eqnarray}
P_E=\frac12 \left[1-\hbox{Erf}
\left(\frac{a}{\sqrt{\kappa}}\right)\right]\stackrel{a\gg1}{\simeq}
\frac{\sqrt{\kappa}}{2a\sqrt{\pi}}\: \exp\left\{-\frac{a^2}{\kappa}\right\}
\:.
\end{eqnarray}
Bob uses a scheme based on heterodyne detection and a threshold strategy as
follows: suppose again that $z_0$ and $z_1$ are real amplitude given by $z_0=-a$ and
$z_1=a$. After having revealed the outcome $z$ 
from the heterodyne detector we employ the following inference rule:
if  Re[z] $<$ 0 then infer bit "0", "1" otherwise. The corresponding error 
probability is given by 
\begin{eqnarray}
P_E^h = \frac12 \Big[p(\hbox{Re}[z]>0|-a)+p(\hbox{Re}[z]<0|a)\Big] 
= \int_{\hbox{Re}[z]<0}\!\!\!\!\!\! 
d^2z \: |{}_1\<\<z|a\>\>_x|^2 = \frac12 \left[1-\hbox{Erf}\left(
\frac{a}{\sqrt{2\sigma_x^2}}
\right)\right]\:,
\end{eqnarray}
where  $\sigma_x^2=1/2 (1-x)(1+x)$. For large $a$ we have $P_E^h\simeq
\sqrt{2\sigma_x^2/a^2\pi}\:\exp\{-a^2/2\sigma_x^2\}$. The error probability of
this Bob' feasible strategy is smaller than optimal Eve's one as far as 
$2\sigma_x^2<\kappa$. The corresponding error probability for Bob' ideal scheme 
is given by $P_E=1/4 \exp\{-4a^2 (1+N)\}$, whereas the analogue "not entangled" 
channel would achieve only $P_E=1/2 [1-\hbox{Erf}(a)]$.
\subsection{Complex alphabet quantum communication}
In this case Alice send through the transmission line the symbol $z_0$, 
chosen from a complex alphabet, encoded into the state $|z_0\rangle\rangle_x$ 
and protected by applying a random displacement $D(\alpha)$ whose amplitude 
is known to Bob. Bob should estimate $z_0$ on the state $|z\rangle\rangle_x$
whereas Eve, for the same task, has at disposal the state 
$D(z) \nu D^\dag (z)$ with
$\nu =  \int d^2\alpha\: g_\kappa(|\alpha|^2)\: |\alpha\>\>_1{}_1\<\<\alpha|$.
If both use heterodyne detection {\em i.e.} the POVM 
$\Pi(z)=|z\>\>_1{}_1\<\<z|$ we have 
\begin{eqnarray}
p_B(z)&=&|{}_1\<\<z|z_0\>\>_x|^2=  \frac{1}{\pi\Delta_x^2}
\exp\left\{-\frac{|z-z_0|^2}{\Delta^2_x}\right\}\nonumber \\ 
p_E(z)&=&{}_1\<\<z| \zeta |z\>\>_1 = 
\int \frac{d^2\alpha}{\kappa\pi} \:e^{-|\alpha|^2/\kappa}\: 
|{}_1\<\<z|D(\alpha)|z_0\>\>_x|^2 = 
\frac{1}{\pi(\Delta_x^2 + \kappa)}
\exp\left\{-\frac{|z-z_0|^2}{\Delta^2_x +\kappa}\right\}\:, 
\end{eqnarray}
and again the security of the protocol is assured by the random 
distribution of the displacing amplitudes.
\section{Degradation of entanglement in active fibers}\label{s:dent}
In applications such teleportation or cryptography one needs to transfer
entanglement among distant partners, and therefore to transmit
entangled states along some kind of channel. For optical implementation 
this is usually accomplished by means of (active) optical fibers. As a 
matter of fact, the propagation of twin-beam in optical fibers 
unavoidably lead to degradation of entanglement due to decoherence 
induced by losses and noise. In this section, we study the evolution of
twin-beam in active optical media, such the pair of optical fibers
that may be used to transmit twin-beam, and analyze the separability 
of the evolved state as a function of the fiber parameters. 
A threshold value for the interaction time, above which the 
entanglement is destroyed, will be analytically derived.  \\ 
If the twin-beam are produced from the 
vacuum by a parametric optical amplifier with evolution operator 
$U =\exp\left[r_0\left(a^\dag b^\dag -ab\right)\right]$, then 
we have $x=\tanh r_0$, whereas the number of photons of the 
twin-beam is $N=2\sinh^2 r_0=2x^2/(1-x^2)$. 
The propagation inside the fibers can be modeled as the coupling of 
each part of the twin-beam with a non zero temperature reservoir. 
The fibers dynamics can be described in terms of the two-mode 
Master equation $
\dot{\varrho_t} \equiv {\cal L} \varrho_t = 
\Gamma_a (1+M_a) L[a] \varrho_t + \Gamma_b (1+M_b) L[b] \varrho_t  + 
\Gamma_a M_a L[a^\dag ] \varrho_t + \Gamma_b  M_b  L[b^\dag] \varrho_t$  
where $\varrho_t\equiv\varrho (t)$, $\Gamma_a=\Gamma_b=\Gamma$ 
denotes the (equal) damping rate, $M_a=M_b=M$ the number of 
background thermal photons, and $L[ O ]$ is the Lindblad superoperator 
$L[ O ] \varrho_t =  O \varrho_t  O^\dag - \frac{1}{2} O^\dag  O 
\varrho_t - \frac{1}{2} \varrho_t O^\dag O \:.$
The terms proportional to $L[a]$ and  $L[b]$ describe the losses, 
whereas the terms proportional to $L[a^\dag]$ and $L[b^\dag]$ describe 
the linear phase-insensitive amplification process taking place into 
the fibers. Of course, the dynamics inside the two fibers are 
independent on each other. 
The master equation can be transformed into a Fokker-Planck
equation for the two-mode Wigner function $W (x_1,y_1;x_2,y_2)$. 
Using the differential representation of the superoperators 
the corresponding Fokker-Planck equation reads as follows 
$\partial_\tau W_\tau(x_1,y_1;x_2,y_2) = \left[ \frac{1}{8}
\left(\sum_{j=1}^2\partial^2_{x_j x_j} + \partial^2_{y_j y_j}\right) 
+ \frac{\gamma}2 \left(\sum_{j=1}^2 \partial_{x_j} x_j+\partial_{y_j}  y_j 
\right) \right] W_\tau(x_1,y_1;x_2,y_2)$, 
where $\tau$ denotes the rescaled time $\tau=\Gamma/\gamma\:t$, 
and the drift term $\gamma$ is given by $\gamma= (2M+1)^{-1}$. 
The Wigner function of a twin-beam is given by
$$ W_0(x_1,y_1;x_2,y_2)=\left(2\pi \sigma_+^2 \: 2\pi \sigma_-^2\right)^{-1}\:
\exp\left[ -\frac{(x_1+x_2)^2}{4\sigma_+^2}
-\frac{(y_1+y_2)^2}{4\sigma_-^2} -\frac{(x_1-x_2)^2}{4\sigma_-^2}
-\frac{(y_1-y_2)^2}{4\sigma_+^2} \right]\:, $$
where  $\sigma^2_+=1/4\exp\{2r_0\}$, $\sigma^2_-=1/4\exp\{-2r_0\}$.
The Gaussian form of the Wigner function is mantained during the evolution
whereas the variances are increased to 
$$\Sigma_+^2 = \left(e^{-\gamma\tau} \sigma_+^2+D^2\right) \qquad 
\Sigma_-^2 = \left(e^{-\gamma\tau} \sigma_-^2+D^2\right),$$ with 
$D^2=\frac{1}{4\gamma }(1 -e^{-\gamma \tau})$.
A necessary condition for disentanglement, or separability, is the positivity 
of the density matrix $\varrho^T$, obtained by partial transposition of the 
original density matrix (PPT condition) \cite{peres}. In general, PPT has been proved 
to be only a necessary condition for separability. However, for some specific 
sets of states PPT is also a sufficient condition. These includes 
Gaussian states (states with a Gaussian Wigner function) of a bipartite 
continuos variable system \cite{simon,duan}. Our analysis is based on this results. 
In fact, the Wigner function of a twin-beam  
is Gaussian, and the evolution in an active medium preserves such 
Gaussian character. Therefore, we are able to characterize the entanglement 
at any time and to give conditions on the parameters to preserve entanglement 
after a given interaction lenght. The PPT condition on the density matrix can be 
rephrased as a condition on the covariance matrix of the Wigner function of the two
modes. In the case of an evolved twin-beam we have that 
the state is separable iff {\em both} the variances satisfies the
condition 
$\Sigma_+^2 \geq \frac14, \; \Sigma_-^2 \geq \frac14.$
Given the parameters $M$, $\Gamma$ and $\lambda$ the threshold 
value $\tau_s$ above which the state become separable is given by
$$ \tau_s=\frac1{\gamma}\: \log \left(1+ \gamma \frac{1-e^{-2\lambda}}{1
-\gamma}\right)=(2M+1)\: 
\log\left(1-\frac{N-\sqrt{N(N+2)}}{2M}\right)\:,$$
(remind that $N$ is the mean photon number of the twin-beam).
In terms of the unrescaled time $t$ the threshold for separability 
reads as 
\begin{eqnarray} t_s=\frac1{\Gamma}\: 
\log\left(1-\frac{N-\sqrt{N(N+2)}}{2M}\right)
\label{thre}\;,
\end{eqnarray}
apart from the case $M=0$ in which the threshold diverges.
Eq. (\ref{thre}) says if the state was initially sufficiently 
entangled the interaction with the environment is not destroying 
its character. In this case we have approximately 
$t_s \simeq \frac{1}{\Gamma} \log(1+\frac{1}{2M})$ 
\section{Summary}\label{s:outro}
The technology of entanglement can be of great help in improving precision, stability 
and performances of quantum optical schemes meant to process quantum
information. In this paper we reviewed some applications of continuous variables 
entangled states, aimed to improve quantum measurements, interferometry and 
communication. Since the optical implementation of quantum information processing 
will involve optical fibers to establish an entangled channel between two
distant users, we also study the evolution of entangled twin-beam of light in an 
active optical medium, such to evaluate the degradation rate of entanglement,
and establish a threshold on the interaction time, above which the entangled
is no longer present and the channel become useless.
\section*{Acknowledgments} This work has been sponsored by the INFM through 
the project PRA-CLON, and by EEC through the projects IST-2000-29681 (ATESIT), 
IST-1999-11053 (EQUIP). 


\begin{thebibliography}{25}
\bibitem{polz} B. Julsgaard, A. Kozhekin, E.  S. Polzik, LANL ArXive quant-ph/0106057 
\bibitem{pati} See for example, {\em Quantum Information Theory with Continuous 
Variables}, S.L.Braunstein and A.K.Pati Eds (Kluwer, 2002).
\bibitem{furu} A. Furusawa et al, Science {\bf 282}, 706 (1998).
\bibitem{geza} L. Duan, G. Giedke, J. I. Cirac, and P. Zoller, Phys. Rev.
Lett. {\bf 84}, 4002 (2000).  
\bibitem{gran} F. Grosshans, P. Grangier, Phys. Rev. Lett. {\bf 88}, 
057902 (2002) 
\bibitem{fabre} M. I. Kolobov and C. Fabre, Phys. Rev. Lett. {\bf 85}
\bibitem{spectr} B. E. A. Saleh, B. M. Jost, H.-B. Fei, and
M. C. Teich, Phys. Rev. Lett {\bf 80} 3483 (1998)
\bibitem{entdec} G. M. D'Ariano, M. G. A. Paris and P. Perinotti, 
Phys. Rev A {\bf 65} 062106  (2002).
\bibitem{tomoch}  G. M. D'Ariano, and P. Lo Presti,
Phys. Rev. Lett. {\bf 86} 4195 (2001) 
\bibitem{entame} G. M. D'Ariano, P. Lo Presti and M. G. A. Paris, 
Phys. Rev. Lett. {\bf 87}, 270404 (2001).
\bibitem{yuen82} E. Arthurs, M. S. Goodman, Phys. Rev. Lett.
{\bf 60}, 2447 (1988)
\bibitem{simon} R. Simon, Phys. Rev. Lett. {\bf 84} 2726 (2000)
\bibitem{helstrom} C.W.Helstrom, {\em Quantum Detection and Estimation 
Theory} (Academic Press, New York, 1976).
\bibitem{intbin} M. G. A. Paris, Phys. Lett. A {\bf 225}, 23 (1997). 
\bibitem{np} J. Neyman, E. Pearson, Proc. Camb. Phil. Soc. {\bf 29}, 
492 (1933); Phil. Trans. Roy. Soc. London A{\bf 231}, 289 (1933).
\bibitem{parat} K.R. Parthasarathy, Inf. Dim. Anal. Quant.  Prob.
{\bf 2}, 557 (1999). 
\bibitem{bon} C. M. Caves, Phys. Rev. D {\bf 23}, 1693 (1981);
R. S. Bondurant and J. H. Shapiro, Phys. Rev. A {\bf 30}, 2548 (1984).
\bibitem{Yuen} H. Yuen, in {\em Quantum Communication, Measurement, and
Computing}, Eds P. Tombesi and O. Hirota, 
(Kluwer Academic/Plenum Publishers, 2001)
\bibitem{bin} Wang Xiang-bin, LANL arXive quant-ph/0204039.
\bibitem{amplent} G. M. D'Ariano, M. G. A. Paris, M. F. Sacchi, 
F. De Martini, Phys. Rev. A {\bf 61}, 063813 (2000).
\bibitem{peres} A. Peres, Phys. Rev. Lett. {\bf 77}, 1413-1415 (1996) 
\bibitem{duan} Lu-Ming Duan, G. Giedke, J. I. Cirac, and P. Zoller, 
Phys. Rev. Lett. {\bf 84} 2722 (2000)
\end{thebibliography}
\end{document}